\documentclass[11pt, a4paper]{article}
\usepackage{moriond,epsfig}

\def\be{\begin{equation}}
\def\ee{\end{equation}}
\def\bea{\begin{eqnarray}}
\def\eea{\end{eqnarray}}

\def\slashiv#1{#1\llap{\sl/}}
\begin{document}
\begin{titlepage} 
\nopagebreak  
{\flushright{ 
        \begin{minipage}{5cm}
          KA--TP--14--2007  \\       
        \end{minipage}        } 
 
} 
\vfill 
\begin{center} 
{\LARGE \bf 
 \baselineskip 0.9cm 
 Hjj production:\\ Signals and CP measurements
} 
\vskip 1.5cm  
{\large   
G.~Kl\"amke and D.~Zeppenfeld
}   
\vskip .2cm  
{{\it Institut f\"ur Theoretische Physik, 
  Universit\"at Karlsruhe, P.O.Box 6980, 76128 Karlsruhe, Germany}
} 
 
 \vskip 
1.3cm     
\end{center} 
 
\nopagebreak 
\begin{abstract}
Higgs boson production in association with two tagging jets will be
mediated by electroweak vector boson fusion and by gluon fusion. For the
gluon fusion process, analysis of the azimuthal angle correlations of the two jets provides
for a direct measurement of the CP-nature of the $Htt$ Yukawa coupling
which is responsible for the effective $Hgg$ vertex.
\end{abstract} 
\vfill 
\end{titlepage} 
\newpage               
%
%
\vspace*{4cm}
\title{HJJ PRODUCTION: SIGNALS AND CP MEASUREMENTS}

\author{ GUNNAR KL\"AMKE and DIETER ZEPPENFELD }
\address{Institut f\"ur Theoretische Physik, 
  Universit\"at Karlsruhe, P.O.Box 6980, 76128 Karlsruhe, Germany}

\maketitle\abstracts{ Higgs boson production in association with two tagging jets will be
mediated by electroweak vector boson fusion and by gluon fusion. For the
gluon fusion process, analysis of the azimuthal angle correlations of the two jets provides
for a direct measurement of the CP-nature of the $Htt$ Yukawa coupling
which is responsible for the effective $Hgg$ vertex.
 } %

\section{Introduction}

Higgs boson production in association with two jets has emerged as a 
promising channel for Higgs boson discovery and for the study of Higgs
boson properties at the LHC. Interest has concentrated on 
vector-boson-fusion (VBF), i.e. the weak process $qq\to qqH$ which is 
mediated by $t$-channel exchange of a $W$ or $Z$, with the Higgs boson 
being radiated off this weak boson. The VBF production cross section measures
the strength of the $WWH$ and $ZZH$ couplings, which, at tree
level, require a vacuum expectation value for the scalar field. Hence the 
VBF channel is a sensitive probe of the Higgs mechanism as the source
of electroweak symmetry breaking.

Another prominent source of $Hjj$ events are second order real emission 
corrections to the gluon fusion process. Such corrections were first 
considered in Ref.~\cite{Kauffman:1996ix,Kauffman:1998yg} in the large 
top mass limit and have subsequently been evaluated for arbitrary quark masses
in the loops which induce the effective coupling of the Higgs
boson to gluons.\cite{DelDuca:2001eu}
For a SM Higgs boson, the generic $Hjj$ cross section from gluon fusion
can somewhat exceed the VBF cross section of a few
pb~\cite{DelDuca:2001eu} and, thus, gluon fusion induced $Hjj$ events 
should also provide useful information on Higgs boson properties. 

\begin{figure}[htb]
\begin{center}
\begin{tabular}{ccc}
\includegraphics[scale=0.68]{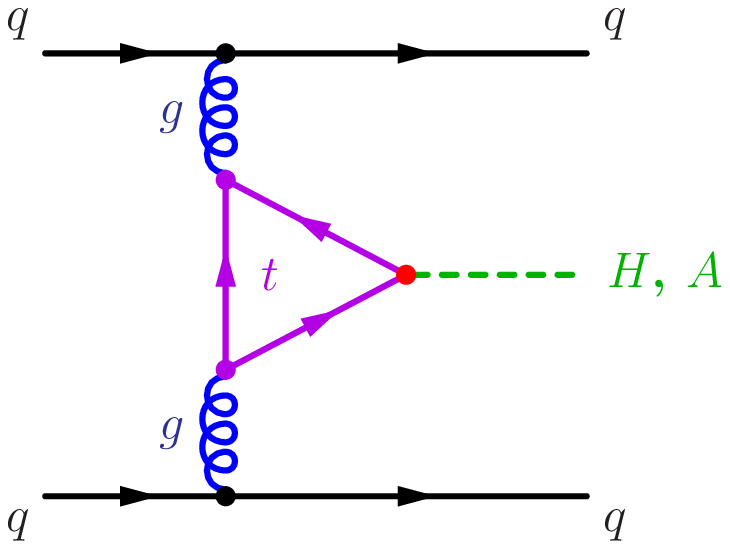}&
\includegraphics[scale=0.327]{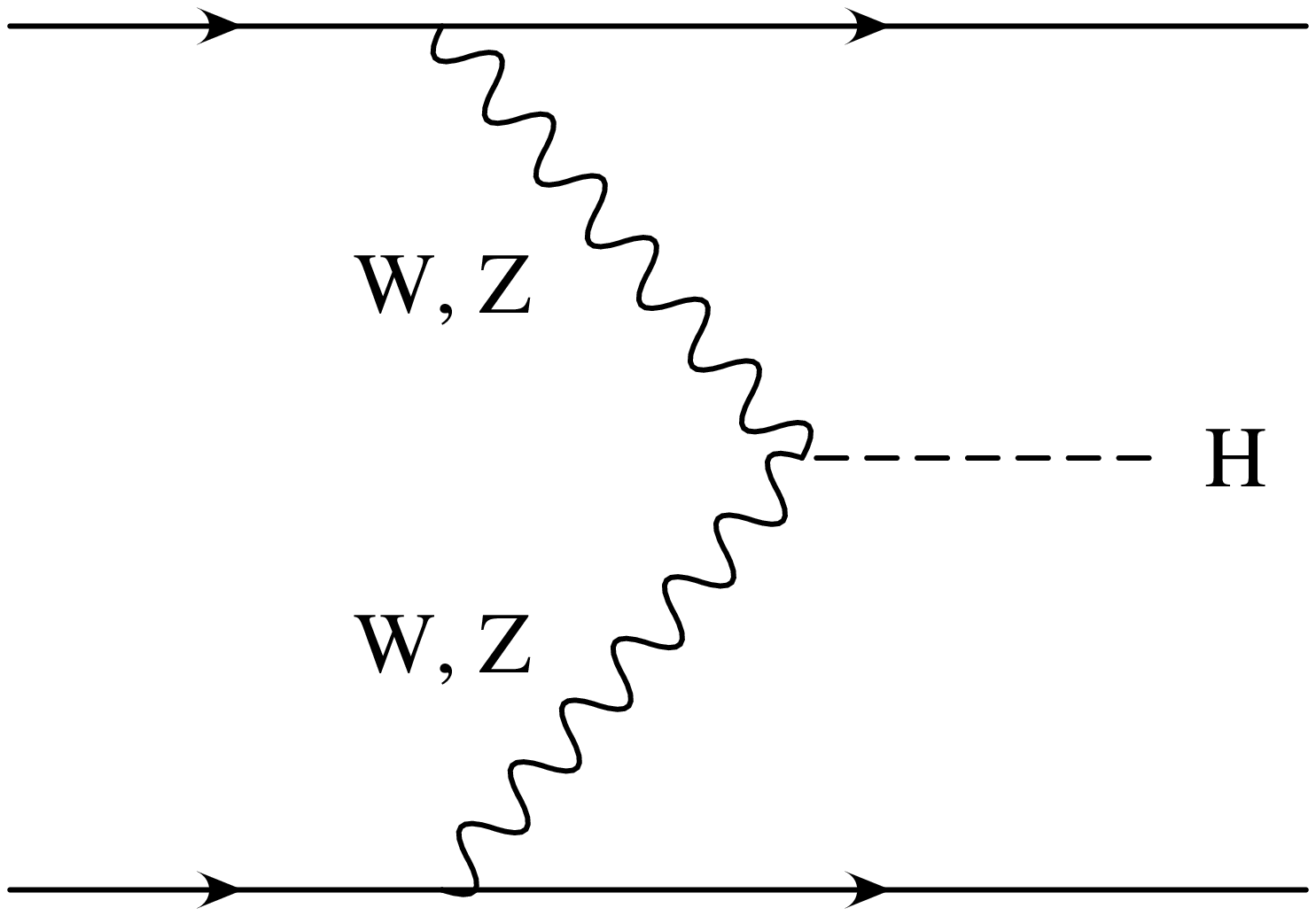}\\
\end{tabular}
\end{center}
\caption[]{\label{fig:tmunu} Quark scattering processes contributing to
  $Hjj$ production via gluon fusion and vector boson fusion. }
\end{figure}

In this contribution we focus on the CP properties of the Higgs  
Yukawa coupling to the top quark, which is given by
\begin{equation}
{\cal L}_Y = y_t H\bar t t +i \tilde y_t A\bar
t\gamma_5 t,
\label{eq:Yuk}
\end{equation}
where $H$ and $A$ denote scalar and  pseudo-scalar Higgs fields. 
Top quark loops
then induce effective couplings of the Higgs boson to gluons which, for
Higgs masses well below $m_t$, can be described by the
effective Lagrangian~\cite{Kauffman:1996ix,Kauffman:1998yg}
\begin{equation}
{\cal L}_{\rm eff} = 
\frac{y_t}{y_t^{SM}}\cdot\frac{\alpha_s}{12\pi v} \cdot H \,G_{\mu\nu}^a\,G^{a\,\mu\nu} +
\frac{\tilde y_t}{y_t^{SM}}\cdot\frac{\alpha_s}{16\pi v} \cdot A \,
G^{a}_{\mu\nu}\,G^{a}_{\rho\sigma}\varepsilon^{\mu\nu\rho\sigma}\;,
\label{eq:ggS}
\end{equation}
where $G^{a}_{\mu\nu}$ denotes the gluon field strength. 
From the effective Lagrangian emerge $Hgg$, $Hggg$ and also $Hgggg$ 
vertices, which correspond to triangle, box and pentagon top quark 
loops and which contribute to gluon fusion processes such as $qq\to
qqH$, $qg\to qgH$ or $gg\to ggH$. One example for the first process and
for the corresponding VBF diagram is shown in  Fig.~\ref{fig:tmunu}.

\section{Azimuthal angle correlations}

Analogous to the corresponding VBF case,\cite{VBF:CP,Hankele:2006ma} 
the distribution of the azimuthal angle between the two
jets in gluon fusion induced $Hjj$ events can be used to determine the
tensor structure of the effective $Hgg$ vertex, which emerges from 
Eq.(\ref{eq:ggS}) as
\begin{equation}
T^{\mu\nu} = a_2\,(q_1\cdot q_2\, g^{\mu\nu} - 
q_1^\nu q_2^\mu) + a_3\, \varepsilon^{\mu\nu\rho\sigma}q_{1\rho}q_{2\sigma}\,,
\label{eq:Tmunu}
\end{equation}
with $a_2 = \frac{y_t}{y_t^{SM}}\cdot\frac{\alpha_s}{3\pi v}$ and 
$a_3 = -\frac{\tilde y_t}{y_t^{SM}}\cdot\frac{\alpha_s}{2\pi v}$.
We assume SM-size couplings in our analysis below.

\begin{figure}[thb]
\centerline{
\includegraphics[scale=0.5]{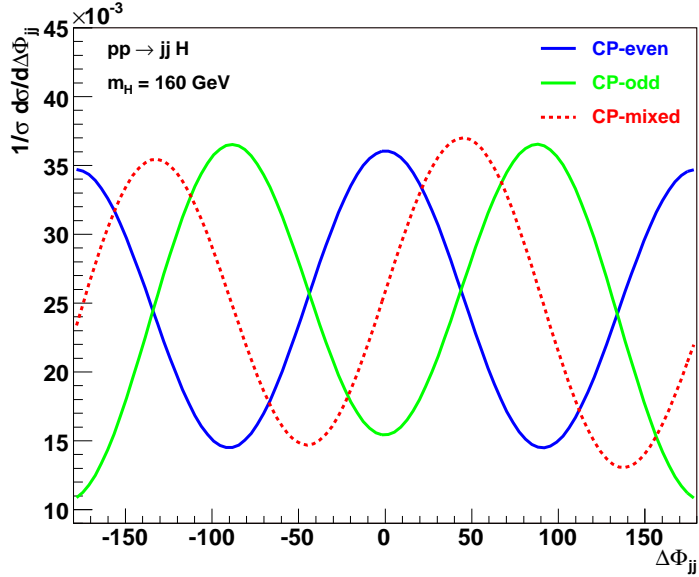}
}
\caption[]{\label{fig:phi1} Normalized distributions of the jet-jet
  azimuthal angle difference as defined in the text. The
  curves are for the SM CP-even case ($a_{3}=0$), a pure CP-odd
  ($a_{2}=0$) and a CP-mixed case ($a_{2}=a_{3}\ne 0$). 
}
\end{figure}

In resolving interference effects between the CP-even coupling $a_2$ and
the CP-odd coupling $a_3$ it is important to measure the sign of the
azimuthal angle between the jets. Naively one might assume that this
sign cannot be defined unambiguously in $pp$ collisions because an
azimuthal angle switches sign when viewed along the opposite beam
direction. However, in doing so, the ``toward'' and the ``away''
tagging jets also switch place, i.e. one should take into account the
correlation of the tagging jets with the two distinct beam
directions. Defining $\Delta\Phi_{jj}$ as the azimuthal angle of the
``away'' jet minus the azimuthal angle of the ``toward'' jet, a switch
of the two beam directions leaves the sign of $\Delta\Phi_{jj}$
intact.\cite{Hankele:2006ma} 
The corresponding distributions, for two jets with 
\begin{equation}
\label{eq:jincl}
p_{Tj} > 30\, {\rm GeV},\qquad|\eta_j| < 4.5,
\qquad |\eta_{j_1}-\eta_{j_2}| > 3.0\;,
\end{equation}
are shown in Fig.~\ref{fig:phi1} for three scenarios of CP-even 
and CP-odd Higgs couplings. All three cases are well distinguishable.
The maxima in the distributions are directly connected to the size
of the parameters $a_2$ and $a_3$ in Eq.~(\ref{eq:Tmunu}). For
\begin{equation}
a_2 = a\,\cos\alpha\, , \qquad a_3 = a\, \sin \alpha\, ,
\end{equation}
the positions of the maxima are at $\Delta\Phi_{jj}=\alpha$ and
$\Delta\Phi_{jj}=\alpha \pm \pi$.

\section{Observability at the LHC}

The azimuthal angle correlations of the two leading jets in gluon fusion 
are fairly independent of the Higgs boson mass and they do not depend 
on the Higgs decay mode, except via kinematical effects due to cuts 
on the decay products. In order to observe them, however, background 
processes have to be suppressed by a sufficient degree. Clearly, this 
depends on which decay channels are available for the Higgs boson. 
The most promising case is a SM-like Higgs boson of mass around 
$m_H\approx 160$~GeV with decay $H\to W^+W^-\to l^+l^-\slashiv{p}_T$ 
($l=e,\mu$). We here give a brief summary of our findings. Details 
of the parton level simulation are given
Ref.~\cite{Klamke:2007cu} Similar to the analogous $H\to WW$ search 
in VBF,\cite{VBF:H} the dominant backgrounds arise from $t\bar t$ 
production in association with 0, 1, or 2 additional jets and from $WWjj$ 
production at order $\alpha^2\alpha_s^2$ (QCD $WWjj$ production) or
at order $\alpha^4$ (EW $WWjj$ production, which includes $H\to WW$ in 
VBF). 

\begin{table}[tbh]
  \caption{Signal and background cross sections and the expected
  number of events for ${\cal L}_{int}=30\,{\rm fb}^{-1}$ at different
  levels of cuts.}
\begin{center}
\begin{tabular}{|c|c|c|c|c|c|}
 \cline{2-6}
\multicolumn{1}{c|}{} & inclusive cuts & \multicolumn{2}{c|}{selection
  cuts} & \multicolumn{2}{c|}{selection
  cuts \& $\Delta\eta_{jj}>3$ }\\
\hline
process & $\sigma$ [fb] & $\sigma$ [fb] & events / {$30\ ,{\rm fb}^{-1}$}
&  $\sigma$ [fb] & events / {$30\,{\rm fb}^{-1}$}\\ 
\hline\hline
GF $pp \rightarrow H +j j$       & 115 & 31.5 & 945 & 10.6 & 318 \\
\hline
EW $pp \rightarrow W^+W^- +j j$ & 75 & 16.5 & 495 & 13.9 & 417 \\
$pp \rightarrow t \bar{t}$       & 6830 & 23.3 & 699 & 1.5 & 45 \\
$pp \rightarrow t \bar{t}+ j$    & 9520 & 51.1 & 1530 & 13.4 & 402 \\
$pp \rightarrow t \bar{t}+ jj $  & 1680 & 11.2 &  336 & 3.8 & 114 \\
QCD $pp \rightarrow W^+W^- +j j$ & 363  & 11.4 & 342 & 3.0 & 90 \\
\hline
sum of backgrounds             & 18500 & 114  & 3410 & 35.6 & 1070 \\
\hline
\end{tabular}
\end{center}
\label{tab:xs2}
\end{table}

The first column of Table~\ref{tab:xs2} gives the expected LHC cross 
sections for the fairly inclusive cuts of Eq.~\ref{eq:jincl} 
(but $|\eta_{j_1}-\eta_{j_2}| >1$) for the tagging jets, defined 
as the two highest $p_T$ jets in an event, and lepton cuts given by
\begin{equation}
p_{T\ell} > 10\, {\rm GeV},\qquad|\eta_\ell| < 2.5,\qquad 
\Delta R_{j\ell} = \sqrt{(\eta_j-\eta_\ell)^2+(\Phi_j-\Phi_\ell)^2} > 0.7\,.
\label{eqn:incl}
\end{equation}
The large top quark background can be suppressed by a veto on events
with a $b$-quark tag on any observable jet. A characteristic feature of
$H\to WW$ decay is the small angular separation and small invariant mass of
the $l^+l^-$ system,\cite{Dittmar:1996ss} which is exploited by the cuts
\begin{equation}
\Delta R_{\ell\ell} < 1.1\,,\qquad m_{\ell\ell} < 75\,{\rm GeV}\,.
\label{eqn:mll}
\end{equation}
The signal is further enhanced by requiring large lepton transverse 
momentum, $p_{Tl}> 30$~GeV, a transverse mass of the 
dilepton/missing $E_T$ system consistent with the Higgs mass, 
$m_T^{WW}<m_H+10$~GeV and not too small compared to the observed
dilepton mass, $m_{\ell\ell}< 0.44 \cdot m^{WW}_T$, and a significant amount 
of missing $p_T$, $\slashiv{p}_T>30$~GeV. The resulting cross sections
and expected event rates for 30~fb$^{-1}$ are given in the second and third
columns of Table~\ref{tab:xs2}: with 30~fb$^{-1}$ the LHC can establish a 
Higgs signal in gluon fusion with a purely statistical error leading to
a significance of $S/\sqrt{B}=16$. 

\begin{figure}[htbp] 
\centerline{
\includegraphics[scale=0.41]{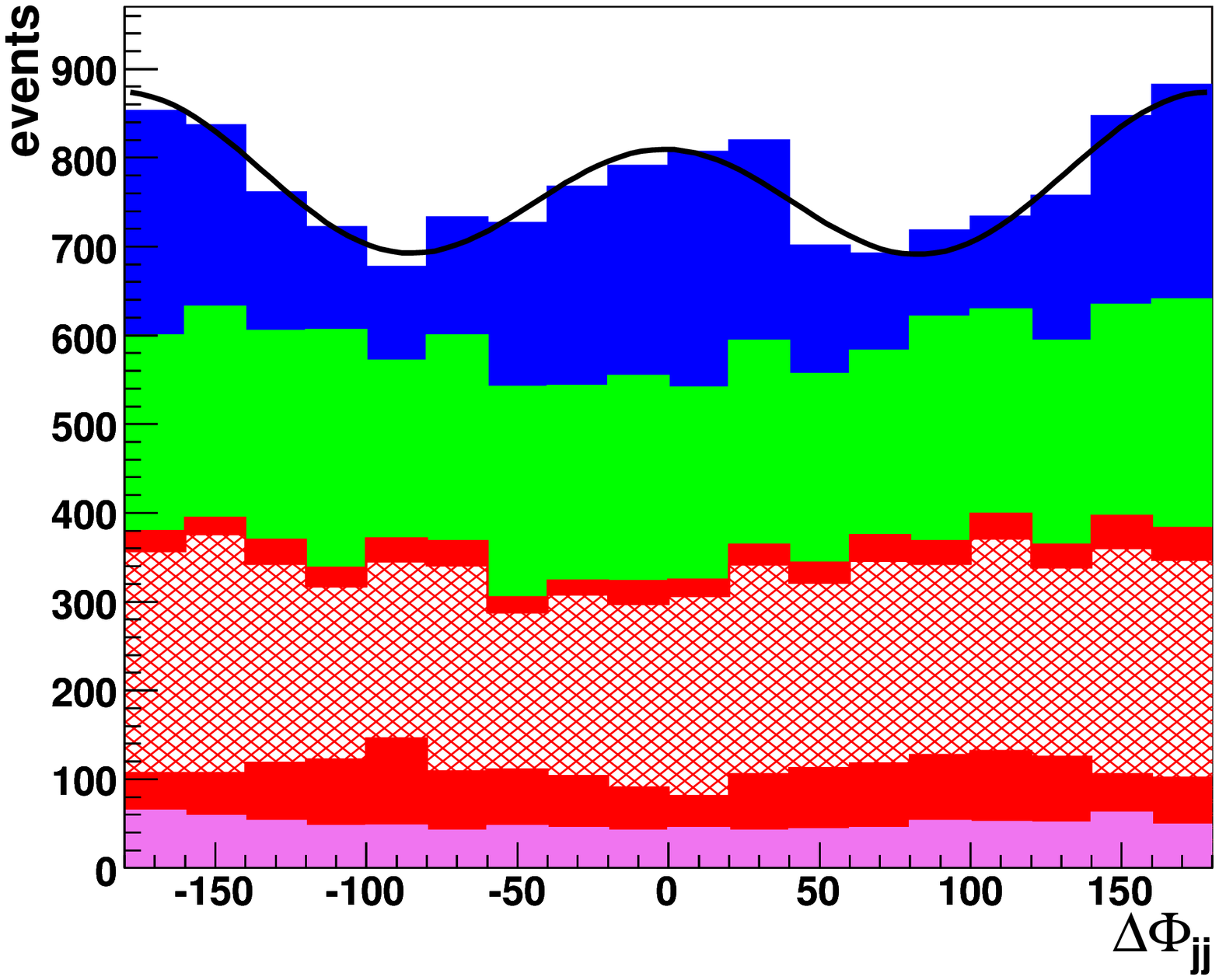}
\includegraphics[scale=0.41]{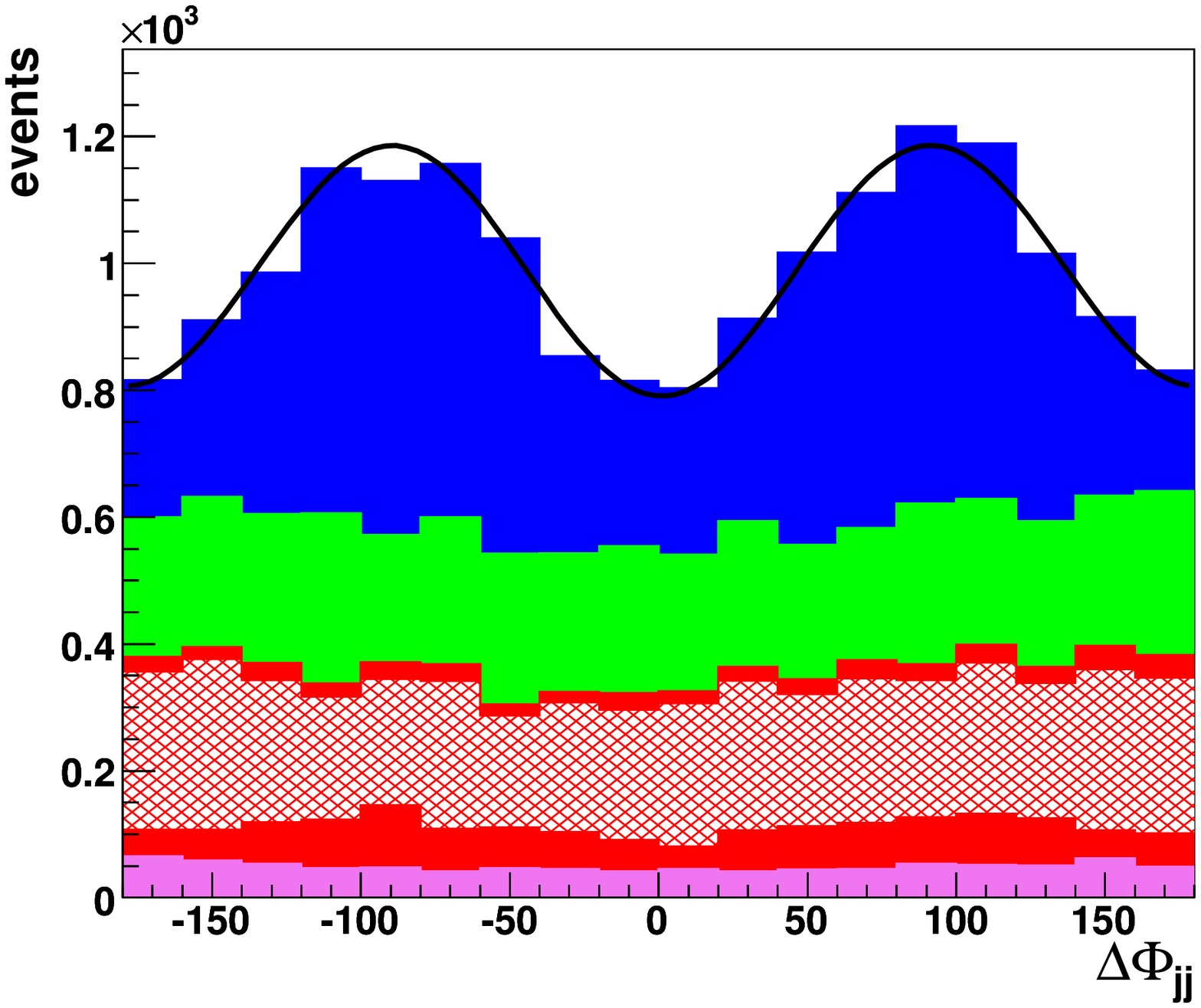}
}
\caption[]{\label{fig:phijjfit} The $\Delta\Phi_{jj}$ distribution for
a pure CP-even coupling {\it (left)} and a pure CP-odd coupling {\it
  (right)} for ${\cal L}_{int} = 300\,{\rm fb}^{-1}$. From top to
bottom: GF signal, EW $W^+W^-jj$, $t\bar{t}$,
$t\bar{t}j$, $t\bar{t}jj$, and QCD $W^+W^-jj$ backgrounds. }
\end{figure}

The resulting event sample of about 950 signal and 3400 background events 
is large enough and sufficiently pure to analyze the azimuthal angle between 
the two tagging jets. One finds, however, that the characteristic modulation
of the $\Delta\phi_{jj}$ distribution in Fig.~\ref{fig:phi1} is most  
pronounced for large rapidity separations of the tagging 
jets.\cite{Klamke:2007cu} Imposing
$\Delta\eta_{jj}>3$, one obtains the cross sections in the second to 
last column of Table~\ref{tab:xs2} and azimuthal angle distributions as 
shown in Fig.~\ref{fig:phijjfit} for an integrated luminosity of 
300~fb$^{-1}$. Already for 30~fb$^{-1}$ of data, however, can one
distinguish the SM expectation in the left panel of Fig.~\ref{fig:phijjfit}
from the CP-odd case in the right panel with a purely statistical power
of more than 5~sigma. We do not expect detector effects or higher order
QCD corrections to substantially degrade these conclusions.

\section*{Acknowledgments}
This research was supported in part by the Deutsche Forschungsgemeinschaft
under SFB TR-9 ``Computational Particle Physics''. G.~K. greatfully 
acknowledges DFG support through the 
Graduiertenkolleg `` High Energy Physics and Particle Astrophysics''.

\end{document}